\documentclass[aps, prd, superscriptaddress, preprintnumbers, floatfix, longbibliography,nofootinbib,twocolumn,10pt]{revtex4-2}

\usepackage[utf8]{inputenc}
\usepackage{newtxtext}
\usepackage[upint]{newtxmath}
\usepackage{microtype}
\usepackage{textcomp}
\usepackage{dsfont}
\usepackage{eucal}
\usepackage{siunitx}
\usepackage{soul}

\usepackage{amsfonts,amsmath,amstext}
\usepackage{graphicx,epstopdf}
\usepackage{float,wrapfig}
\usepackage{subfigure, psfrag}
\usepackage{dsfont,bm}
\usepackage{xcolor}
\usepackage[normalem]{ulem}
\usepackage[colorlinks=true,urlcolor=blue,citecolor=blue,linkcolor=blue]{hyperref}
\hypersetup{breaklinks=true}

\definecolor{DarkRed}{rgb}{0.65,0,0}%
\definecolor{Green}{rgb}{0,0.3,0.3}
\definecolor{Purple}{rgb}{0.3,0,0.65}
\definecolor{Red}{rgb}{1,0,0}
\definecolor{Blue}{rgb}{0,0,0.85}
\definecolor{Magenta}{rgb}{1,0,1}

\definecolor{Red}{rgb}{1,0,0}
\definecolor{Green}{rgb}{0,0.5,0}

\newcommand{\sign}[1]{\,\mbox{sgn}\left({#1}\right)}

\newcommand{\IM}[1]{\,\mbox{Im}\left\{{#1}\right\}}
\newcommand{\df}[1]{\,\delta{\left(#1\right)}}

\begin{document}

\title{Thermoelectric Effect in Altermagnet-Superconductor Junctions}
\date{\today}

\author{Pavlo O. Sukhachov}
\email{pavlo.sukhachov@ntnu.no}
\affiliation{Center for Quantum Spintronics, Department of Physics, Norwegian University of Science and Technology, NO-7491 Trondheim, Norway}

\author{Erik Wegner Hodt}
\affiliation{Center for Quantum Spintronics, Department of Physics, Norwegian University of Science and Technology, NO-7491 Trondheim, Norway}

\author{Jacob Linder}
\email{jacob.linder@ntnu.no}
\affiliation{Center for Quantum Spintronics, Department of Physics, Norwegian University of Science and Technology, NO-7491 Trondheim, Norway}

\begin{abstract}
We propose altermagnet-superconductor junctions as a way to achieve a thermoelectric response in metals free of external or stray magnetic fields. We combine qualitative analysis in a simplified model with a more rigorous approach based on the inverse proximity effect in the functional-integral formulation. We show that coupling an altermagnet to a superconductor in a bilayer induces a momentum-dependent spin-splitting in the superconductor. When tunneling occurs between this bilayer and a different altermagnet, a spin-dependent particle-hole symmetry breakdown arises in the transport, which leads to a thermoelectric response. Our results show that the altermagnet-superconductor junctions may achieve comparable thermoelectric performance to ferromagnet-superconductor junctions, featuring a nonmonotonic dependence of the figure of merit on the strength of the altermagnetic splitting. We also point out an often overlooked fact regarding the inverse proximity effect in superconductors, namely that even in a normal metal-superconductor junction there is a minigap in the superconductor, which gives rise to a four-peak structure in the DOS reminiscent of spin-split superconductors. Our results show that altermagnetic metals, unlike conventional antiferromagnets, can be used for efficient cryogenic thermoelectricity.
\end{abstract}

\maketitle

\section{Introduction}
\label{sec:intro}

Thermoelectric effects play a crucial role in energy harvesting and cooling of electronics. While the corresponding conversion efficiency and figure of merit $ZT$ are usually low compared to conventional heat engines, thermoelectric devices are highly scalable and do not involve moving parts. Therefore, materials with a strong thermoelectric response are expected to be useful in nanodevices.

Due to the tunability of the Fermi level and the possibility of achieving strong particle-hole symmetry breakdown, semiconductors and semimetals demonstrate a large thermoelectric response~\cite{Shakouri:rev-2011}. Conventional superconductors emerging out of a metallic normal state, on the other hand, are on their own poor thermoelectric materials. Despite having a gap, the thermoelectric response of superconductors resembles that of metals rather than semimetals. This is explained by the inherent particle-hole symmetry of superconductors.

The particle-hole symmetry obstruction can be overcome by introducing a spin-splitting field, for instance by coupling a superconductor to a ferromagnet~\cite{kalenkov_prl_12, machon_prl_13, Ozaeta-Heikkila:2014,Giazotto-Bergeret:2015,Linder-Bathen:2016} through the inverse proximity effect \cite{tedrow_prl_71,  tokuyasu_prb_88, hao_prb_90, Bergeret-Martin-Rodero:2005}. Subsequently, a large figure of merit can be obtained by coupling such a superconductor to a spinful system such as a spin-active interface. For example, by considering a junction between a spin-split superconductor to a ferromagnet, the figure of merit $ZT\simeq 4$ was predicted in Ref.~\cite{Ozaeta-Heikkila:2014}, which could be further increased to $ZT\simeq 40$ in junctions involving two spin-split superconductors~\cite{Linder-Bathen:2016}. Such figures of merit exceed typical values for commercial thermoelectric materials featuring $ZT\simeq1$~\cite{Snyder-Toberer:rev-2008}, such as Bi$_2$Te$_3$, as well as the best available thermoelectric material sodium-doped PbTe~\cite{Biswas-Kanatzidis:2012} and thin-film Heusler alloys~\cite{Hinterleitner-Bauer:2019} with $ZT\lesssim 6$. Moreover, since typical thermoelectric materials perform much better at high temperatures (room temperature or higher), superconductor-ferromagnet structures serve the important purpose of providing a large thermoelectric effect in materials at cryogenic temperatures.

The need for magnetic fields or ferromagnets, however, hinders the miniaturization of thermoelectric devices based on superconductors. Indeed, stray magnetic fields are generically undesirable, especially if such thermoelectric devices are combined with other devices based on spintronics, such as magnetoresistive or spin-transfer torque architectures. Therefore, it is imperative to seek new ways to create high-$ZT$ materials at low temperatures without any stray magnetic fields. To avoid using ferromagnets or external magnetic fields, we propose a new class of thermoelectric devices based on superconducting junctions with altermagnets.

Altermagnets is a recently discovered class of materials showing momentum-dependent spin-splitting that is distinct from relativistically spin-orbit coupled systems~\cite{Smejkal-Jungwirth:2022b}. The spin-splitting in altermagnets originates from a crystal lattice geometry and ordering of localized spins which combined break parity and time-reversal symmetry. This lifts spin-degeneracy for the band structure in the Brillouin zone of the itinerant electrons, which interact through regular exchange coupling with the localized spins, except for certain high-symmetry points in momentum space. Unlike ferromagnets, however, the net magnetization of altermagnets is zero after integrating over the Brillouin zone~\footnote{According to the extended classification in Ref.~\cite{Cheong-Huang:2024}, altermagnets with vanishing magnetization belong to types II and III altermagnets.}, which allows for magnetic-field-free devices~\footnote{As we show in Ref.~\cite{Sukhachov-Hodt-Linder:2024}, finite size effects in altermagnets lead to a nonzero magnetization which is concentrated near the boundaries. In this study, we ignore this effect.}.
Altermagnets were predicted via \textit{ab initio} calculations in several material candidates including metals like RuO$_2$~\cite{Ahn-Kunes:2019,Gonzalez-Hernandez-Zelezny:2021,Smejkal-Sinova:2020} and Mn$_5$Si$_3$~\cite{Reichlova-Smejkal:2020}, and semiconductors/insulators like MnTe~\cite{Gonzalez-Hernandez-Zelezny:2021,Smejkal-Jungwirth:2022b}, CrSb~\cite{Smejkal-Jungwirth:2022b}, MnF$_2$~\cite{Yuan-Zunger:2020,Egorov-Evarestov:2021}, and La$_2$CuO$_4$~\cite{Smejkal-Jungwirth:2022b}.
Recent ARPES measurements in MnTe~\cite{Lee-Kim:2023,Krempasky-Jungwirth:2024}, RuO$_2$~\cite{Fedchenko-Elmers:2023,Li-Felser-Ma:2024}, and CrSb~\cite{Reimers-Jourdan:2023} have corroborated several of these predictions. Another piece of evidence supporting altermagnetism is provided by measuring the anomalous Hall effect~\cite{Feng-Liu:2022,Tschirner-Veyrat:2023} and spin-splitting torques~\cite{Bose-Ralph:2022,Bai-Song:2022,Karube-Nitta:2022}.

In this paper, we propose to leverage the momentum-dependent spin splitting of altermagnets to obtain a thermoelectric response in the absence of any magnetization or external magnetic field. Our model setup is shown in Fig.~\ref{fig:model-setup}. We introduce momentum-dependent spin splitting in the superconducting (SC) part of the junction by proximitizing it to an altermagnet (AM). This breaks the particle-hole symmetry for each of the spins. Then, by coupling the AM-SC heterostructure to another altermagnet, we take advantage of directional tunneling through the planar interface between the heterostructure and the altermagnetic contact. Such tunneling allows one to distinguish between the spin-split particles and, therefore, leads to an effectively spin-active interface providing the final ingredient needed to achieve a thermoelectric response without any magnetic fields.

Our paper is organized as follows. In Sec.~\ref{sec:model}, we introduce an effective model and provide a qualitative description of the thermoelectric effect in AM-SC-AM heterostructures. A more rigorous approach to the SC-AM bilayer based on the inverse-proximity effect in the functional-integral approach is provided in Sec.~\ref{sec:path}, which will be shown to largely confirm the results obtained in the effective model. The results are discussed and summarized in Sec.~\ref{sec:Summary}. An additional set of spectral functions and a lattice model are presented in Appendices~\ref{sec:App-1} and \ref{sec:App-2}, respectively. Throughout this paper, we use $\hbar=k_{\rm B}=1$.

\section{Qualitative discussion}
\label{sec:model}

In this Section, we review the thermoelectric response of ferromagnet (FM)-SC heterostructures, introduce an effective model for an AM-SC-AM heterostructure, and analyze the key ingredients needed to achieve the thermoelectric response. We discuss what makes the thermoelectric response of AM-SC-AM heterostructures different from that of their ferromagnetic counterparts.

\subsection{Effective model, setup, and key definitions}
\label{sec:model-setup}

Let us start by defining the Hamiltonians of the constituent parts of heterostructures, namely, ferromagnets, altermagnets, and superconductors. We use the following Hamiltonian for a ferromagnet:
\begin{equation}
\label{proximity-F-SC-HF}
H_{F}(\mathbf{p}) = \xi_p + \sigma_z h.
\end{equation}
Here $\xi_p=p^2/(2m)-\mu$, ${\bf p} = (p_x,p_y)$ is momentum in 2D, $p=|{\bf p}|$ is its magnitude, $m$ is mass, $\mu$ is the Fermi energy, $\sigma_z$ is the Pauli matrix acting in the spin space, and $h$ is the exchange field. We consider here calculations in 2D rather than 3D for simplicity since it is the smallest dimension which allows us to capture the characteristic spin-polarized band structure of altermagnets. The thermoelectric effects to be predicted throughout this manuscript persist also in 3D.

\begin{figure*}[!ht]
\centering
\includegraphics[width=0.45\textwidth]{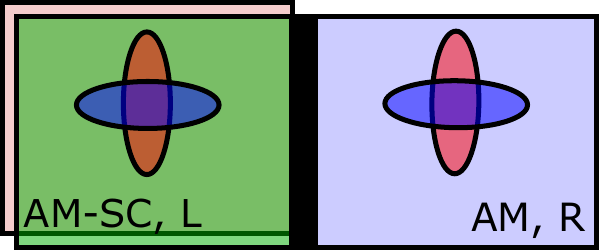}
\caption{
Schematic setup for the AM-SC bilayer that has a planar interface with another AM.
}
\label{fig:model-setup}
\end{figure*}

The Hamiltonian of a $d$-wave altermagnet reads~\cite{Reichlova-Smejkal:2020,Smejkal-Jungwirth:2022}
\begin{eqnarray}
\label{interface-H-R}
H_{AM}(\mathbf{p}) &=& \xi_p + \sigma_z \frac{1}{2m}\left[t_1
\left(p_x^2 -p_y^2\right) +2t_2
p_x p_y\right] \nonumber\\
&=&\xi_p + \sigma_z \left( \xi_p +\mu\right)J_{AM}(\varphi),
\end{eqnarray}
where the dimensionless parameters $t_1$ and $t_2$ determine the orientation and strength of the altermagnetic spin splitting defined via
\begin{eqnarray}
\label{interface-H-JR}
J_{AM}(\varphi) = t_1 \cos{(2\varphi)} +t_2 \sin{(2\varphi)}.
\end{eqnarray}
The Fermi surface in the case $t_1\neq0$ and $t_2=0$ is schematically shown in Fig.~\ref{fig:model-setup}; one should rotate the altermagnetic lobes by $\pi/4$ for $t_1=0$ and $t_2\neq0$.

To develop physical intuition, we first use a simplified model of a superconductor proximitized to a ferromagnet or an altermagnet. In the BdG representation, the Hamiltonian of the proximitized superconductor reads~\footnote{In writing Hamiltonian (\ref{model-setup-H-AM-SC}), we use the following representation of the Nambu spinor: $\Psi_{N} = \left\{\hat{a}_{\mathbf{p},\uparrow}, \hat{a}_{\mathbf{p},\downarrow}, \hat{a}_{-\mathbf{p},\downarrow}^{\dag}, -\hat{a}_{-\mathbf{p},\uparrow}^{\dag}\right\}$ with $\hat{a}_{\mathbf{p},s}^{\dag}$ and $\hat{a}_{\mathbf{p},s}$ being the fermion creation and annihilation operators with the spin projection $s$.}
\begin{equation}
\label{model-setup-H-AM-SC}
\hat{H}_{SC}(\mathbf{p}) =
    \begin{pmatrix}
    \xi_p +\sigma_z h(\mathbf{p}) & \Delta\\
    \Delta & -\xi_p +\sigma_z h(-\mathbf{p})
    \end{pmatrix},
\end{equation}
where $\Delta$ is the spin-singlet $s$-wave superconducting gap. If the superconductor is proximitized by a ferromagnet, the exchange field is momentum-independent $h(\mathbf{p}) = h$. On the other hand, the inverse proximity effect to an altermagnet can be effectively described as a momentum-dependent exchange field $h(\mathbf{p}) = (\xi_p +\mu) J_{AM}'(\varphi)$, where, in general, $J_{AM}'(\varphi)\neq J_{AM}(\varphi)$. We justify this model in Sec.~\ref{sec:path} by using the functional-integral approach.

The schematic setup of an FM-SC-FM or AM-SC-AM heterostructure that can be used to observe thermoelectric effects is shown in Fig.~\ref{fig:model-setup}. The proximitized superconductor is on the left-hand side of the junction (i.e., the FM-SC or AM-SC part) and the ferromagnetic or altermagnetic contact is on the right-hand side. The contacts are separated by a tunneling barrier. In what follows, we will use the subscripts and superscripts $L$ and $R$ to distinguish the left and right parts of the junction.

To calculate the thermoelectric response, we use the standard tunneling Hamiltonian approach~\cite{Levitov-Shytov:book,Schrieffer:book,Mahan:book-2013}. We assume that the insulating barrier is high enough to justify the perturbative treatment in the tunneling coefficient. Then, for weak tunneling, we can neglect the contribution of Andreev processes and use the following expression for the spin-resolved tunneling current~\cite{Levitov-Shytov:book,Schrieffer:book,Mahan:book-2013}:
\begin{widetext}
\begin{eqnarray}
\label{model-setup-Is}
I_{s}(V) &=& 4e\pi^3 \int_{-\infty}^{\infty} d\omega \sum_{\mathbf{k},\mathbf{p}} |t_{\mathbf{p},\mathbf{k};s}^{(R)}|^2 \sum_{\eta=\pm} 
\eta \IM{G_{L;\eta,s}(\omega;\mathbf{p})} \IM{G_{R;\eta,s}(\omega -\eta eV;\mathbf{k})}\left[f_{R}(\omega-\eta eV) -f_{L}(\omega)\right] \nonumber\\
&\approx& 4e \pi^3 \int_{-\infty}^{\infty} d\omega \sum_{\mathbf{k},\mathbf{p}} |t_{\mathbf{p},\mathbf{k};s}^{(R)}|^2 \sum_{\eta=\pm} 
\IM{G_{L;\eta,s}(\omega;\mathbf{p})} \IM{G_{R;\eta,s}(\omega;\mathbf{k})} \frac{eV}{4T\cosh^2{\left(\frac{\omega}{2T}\right)}},
\end{eqnarray}
\end{widetext}
where $s=\pm$ is the spin projection, $\eta=\pm$ correspond to the particle-hole degree of freedom, $-e$ is the electron's charge, $V$ is the bias voltage between the right and left contacts in the heterostructure, $f_{L/R}(\omega)$ is the Fermi-Dirac distribution function, and $T$ is temperature. In the last expression in Eq.~(\ref{model-setup-Is}), we expanded in small $|eV|/\Delta$ in the second line. As one can see, the current (\ref{model-setup-Is}) is determined by the tunneling coefficient, the overlap of the spectral functions, and the difference in the occupation numbers.

The tunneling coefficient for a planar interface preserves the momentum components parallel to the interface and is $|t_{\mathbf{p},\mathbf{k};s}^{(R)}|^2 = [t_s^{(R)}(\varphi)]^2 \delta_{p_{\parallel},k_{\parallel}}/L$ with $L$ being the size of the interface. Here, the angular dependence of $t_s^{(R)}(\varphi)$ represents the fact that tunneling normal to the interface has the largest probability. Furthermore, the tunneling coefficient may in general depend on the spin projection; the dependence is crucial for the models of the FM-SC-FM heterostructure in Refs.~\cite{Ozaeta-Heikkila:2014,Linder-Bathen:2016}.

\begin{figure}[!ht]
\centering
\includegraphics[width=0.42\textwidth]{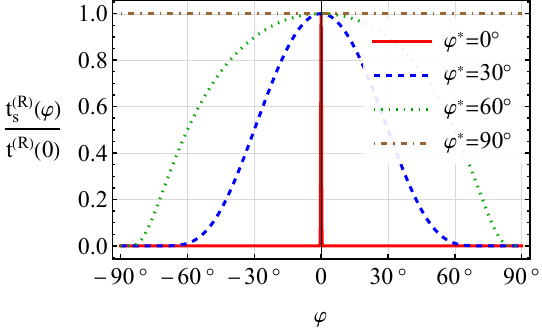}
\caption{
The angular dependence of the normalized tunneling coefficient $t_s^{(R)}(\varphi)/t^{(R)}(0)$ at $P=0$, see Eq.~(\ref{model-setup-t-phi}) for its definition.
}
\label{fig:t-phi}
\end{figure}

We model $t_s^{(R)}(\varphi)$ as
\begin{equation}
\label{model-setup-t-phi}
t_s^{(R)}(\varphi) = \left(1+\eta s P\right) t^{(R)}(0) \left(\frac{1}{2}\right)^{\frac{\cos{\varphi^*}}{\cos{\varphi}} \frac{1-\cos{\varphi}}{1-\cos{\varphi^*}}},
\end{equation}
where we defined the tunneling angle $\varphi^*$ as the angle at which $t_s^{(R)}(\varphi^*)=t_s^{(R)}(0)/2$ and $|P|<1$ is the spin polarization of the interface.
We show the angular dependence of the tunneling function (\ref{model-setup-t-phi}) in Fig.~\ref{fig:t-phi}.

In the case of the AM-SC-AM junction, the retarded Green's functions used in Eq.~(\ref{model-setup-Is}) are
\begin{eqnarray}
\label{toy-model-G-R}
G_{R;\eta,s}(\omega,\mathbf{p}) &=& \eta \frac{1}{\eta (\omega +i 0^{+}) -\xi_{p} - \eta s \left(\xi_{p}+\mu\right) J_R(\varphi)},\\
\label{toy-model-G-L}
G_{L;\eta,s}(\omega,\mathbf{p}) &=& \eta \frac{\omega_{\eta,p,s} +\xi_{p}}{\omega_{\eta,p,s}^2 -\xi_{p}^2 -\Delta^2},
\end{eqnarray}
where $\omega_{\eta,p,s} = \eta (\omega +i 0^{+})  -\eta s (\xi_p +\mu)J_L(\varphi)$. Similar Green's functions albeit with $(\xi_p +\mu)J_{R/L}(\varphi) \to h_{L/R}$ are used for the FM-SC-FM junction.

In the linear response regime, the electric $I(V) = \sum_{s}I_{s}(V)$ and heat $\dot{Q}(V)  = \sum_{s}\dot{Q}_{s}(V)$ currents are conveniently defined as
\begin{equation}
\label{toy-model-momentum-I-Q-L}
\begin{pmatrix}
I(V)\\
\dot{Q}(V)
\end{pmatrix} =
\begin{pmatrix}
L_{11} & L_{12}\\
L_{21} & L_{22}
\end{pmatrix} \begin{pmatrix}
V\\
\frac{\delta T}{T}
\end{pmatrix},
\end{equation}
where $\delta T = T_R - T_L \ll T$ is the temperature difference. The Onsager reciprocal relations require $L_{12}=L_{21}$. As follows from Eq.~(\ref{model-setup-Is}), the Onsager coefficients are
\begin{eqnarray}
\label{toy-model-momentum-Lij-def}
L_{ij} &=& 4\pi^3 \int_{-\infty}^{\infty} d\omega \sum_{\mathbf{k},\mathbf{p}} |t_{\mathbf{p},\mathbf{k}}^{(R)}|^2 \sum_{\eta,s} \IM{G_{L;\eta,s}(\omega;\mathbf{p})} \nonumber\\
&\times&\IM{G_{R;\eta,s}(\omega;\mathbf{k})} \frac{F_{ij;\eta}(\omega)}{4T\cosh^2{\left(\frac{\omega}{2T}\right)}},
\end{eqnarray}
where $F_{11;\eta}(\omega)=e^2$, $F_{12;\eta}(\omega)=\eta e \omega$, and $F_{22;\eta}(\omega)=\omega^2$.

To compare the strength of the thermoelectric response in superconducting heterostructures with other thermoelectric materials, we use the Seebeck coefficient and the figure of merit
\begin{eqnarray}
\label{toy-model-momentum-S-def}
S &=& -\frac{L_{12}}{L_{11} T},\\
\label{toy-model-momentum-ZT-def}
ZT &=& \left(\frac{L_{11} L_{22}}{L_{12}^2} -1\right)^{-1},
\end{eqnarray}
respectively. The Seebeck coefficient $S$ or thermopower is defined as a voltage due to a temperature difference in the open circuit. The figure of merit $ZT$ characterizes the power conversion efficiency; the system reaches the Chambadal–Novikov~\cite{Novikov:1957,Chambadal:1957,Novikov:1958} efficiency $\eta =1 -\sqrt{T_{\rm cold}/T_{\rm hot}}$ at $ZT\to\infty$.

In what follows, we analyze the Onsager coefficients and present the Seebeck coefficient and the figure of merit for a well-studied case of FM-SC-FM heterostructures. These results will be contrasted with the thermoelectric response of the AM-SC-AM heterostructure in Sec.~\ref{sec:model-AM}.

\subsection{FM-SC-FM heterostructure}
\label{sec:model-FM}

In the case of the FM-SC-FM heterostructure, the Onsager coefficient (\ref{toy-model-momentum-Lij-def}) reads
\begin{widetext}
\begin{eqnarray}
\label{toy-model-FM-Lij}
L_{ij} &=& 2\pi^2 \sqrt{2m} L^3 \nu_0\int_{-\infty}^{\infty} d\omega \int_{-\infty}^{\infty} d\xi_p \int_0^{2\pi}\frac{d\varphi}{2\pi} \sum_{\eta,s} \eta |t_s^{(R)}(\varphi)|^2\frac{\Theta{\left(\eta(\omega -sh_R) +\mu -(\xi_p+\mu) \sin^2{\varphi}\right)}}{\sqrt{\eta(\omega -sh_R) +\mu -(\xi_p+\mu) \sin^2{\varphi}}} \nonumber\\
&\times&\sign{\omega -s h_L} \left\{\eta \left[\omega -s h_L\right] +\xi_p\right\} \df{\left[\omega -s h_L\right]^2 -\xi_p^2 -\Delta^2} \frac{F_{ij;\eta}(\omega)}{4T\cosh^2{\left(\frac{\omega}{2T}\right)}},
\end{eqnarray}
\end{widetext}
where we used Eqs.~(\ref{toy-model-G-R}) and (\ref{toy-model-G-L}) with $(\xi_p +\mu)J_{R/L}(\varphi) \to h_{L/R}$, $\nu_0=m/(2\pi)$ is the normal-state density of states (DOS), and $\Theta{(x)}$ is the unit step function.

If we assume normal tunneling, i.e., $\varphi^*\to0$ in Eq.~(\ref{model-setup-t-phi}), the expression for the Onsager coefficient (\ref{toy-model-FM-Lij}) simplifies
\begin{widetext}
\begin{equation}
\label{toy-model-FM-Lij-1}
L_{ij} \stackrel{\varphi^*\to0}{\approx}
2\pi^2 \sqrt{2m} L^3 \nu_0\int_{-\infty}^{\infty} d\omega  \sum_{\eta,s} |t^{(R)}(0)|^2  \frac{\left(1+\eta s P\right)^2}{\sqrt{\mu -\eta sh_R}} \frac{\left|\omega -s h_L\right|}{\sqrt{\left(\omega -s h_L\right)^2-\Delta^2}} \Theta{\left(\left(\omega -s h_L\right)^2-\Delta^2\right)} \frac{F_{ij;\eta}(\omega)}{4T\cosh^2{\left(\frac{\omega}{2T}\right)}},
\end{equation}
\end{widetext}
where we also assumed that $\mu\gg \Delta, T, h_L$ but $\mu \gtrsim h_R$. As one can see from the above expression, the combination of the spin-dependent tunneling quantified by $P$ or a strong exchange field $h_R \sim \mu$~\footnote{The spin-dependent tunneling and the exchange field $h_R$ play a similar role in creating the asymmetry between spin projections.} and the spin-split DOS in the superconductor quantified by $h_L$ allows for a nontrivial $L_{12}$. Without both of these ingredients, $L_{12}=0$. Indeed, the exchange field in the superconductor breaks the particle-hole symmetry separately for each of the spin projections; the symmetry is restored after summing over all spins. By introducing spin-dependent tunneling or a strong exchange field in the ferromagnetic contact, we create the asymmetry between the spin projections, hence, allowing for the particle-hole symmetry breakdown and thermoelectric response.

\begin{figure*}[!ht]
\centering
\subfigure[]{\includegraphics[width=0.31\textwidth]{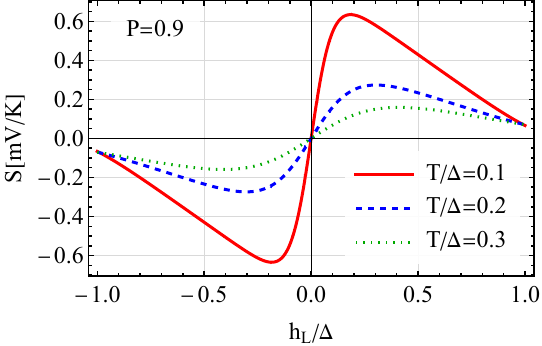}}
\subfigure[]{\includegraphics[width=0.31\textwidth]{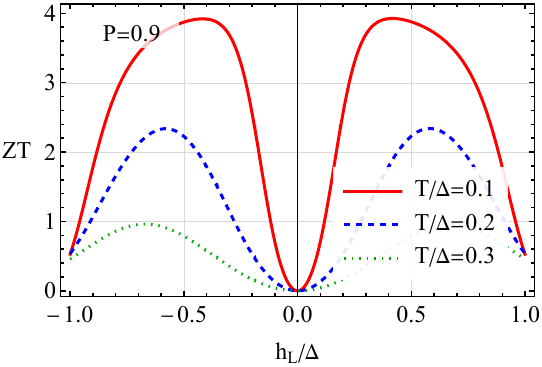}}
\subfigure[]{\includegraphics[width=0.31\textwidth]{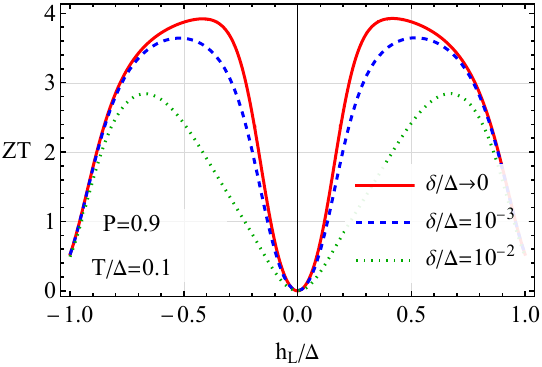}}
\caption{
The Seebeck coefficient $S$ (panel (a)) and the figure of merit $ZT$ (panel (b)) as a function of the exchange field amplitude $h_{L}$ in the FM-SC-FM heterostructure for a few values of $T$. The figure of merit for a few values of the broadening parameter $\delta$ is shown in panel (c). The spin-filtering coefficient is $P=0.9$ and we assume normal tunneling $\varphi^{*}\to0$, see Eqs.~ (\ref{toy-model-momentum-S-def}), (\ref{toy-model-momentum-ZT-def}), and (\ref{toy-model-FM-Lij-1}), for the definition of the Seebeck coefficient, the figure of merit, and the transport coefficients, respectively.
}
\label{fig:toy-model-FM-S-ZT}
\end{figure*}

We present the Seebek coefficient $S$ and the figure of merit $ZT$ for the FM-SC-FM heterostructure in Fig.~\ref{fig:toy-model-FM-S-ZT} assuming $\mu \gg h_R$. The obtained results agree with those in Ref.~\cite{Ozaeta-Heikkila:2014} if one ignores the dependence of the superconducting gap on the magnetic field. As one can see, the figure of merit is nonmonotonic and reaches $4$ at $P=0.9$, which exceeds typical thermoelectric materials~\cite{Snyder-Toberer:rev-2008}. In the limit of low temperature, $\lim_{T\to0} ZT\to P^2/\left(1-P^2\right)$. We note, however, that this limiting value is not achievable in disordered systems. Indeed, by introducing the finite broadening $i0^{+}\to i\delta$ in the FM-SC Green's function (\ref{toy-model-G-L}), we found that the figure of merit, in particular, at low temperatures, is suppressed. The suppression is illustrated in Fig.~\ref{fig:toy-model-FM-S-ZT}(c). In passing, we note that the assumption of normal tunneling is not crucial and can be relaxed. Nonzero tunneling angles introduce only quantitative correction leading to the suppression of thermoelectric effects.

\subsection{AM-SC-AM heterostructure}
\label{sec:model-AM}

In this Section, we address the thermoelectric response of the AM-SC-AM heterostructure by using an effective model with a momentum-dependent exchange field, see Fig.~\ref{fig:model-setup} for the setup. Before proceeding to the calculations, let us discuss what makes altermagnetic heterostructures different from their ferromagnetic counterparts. First of all, since the total magnetization of altermagnets vanishes, there is no spin-splitting of the DOS neither in the proximitized superconductor nor in the altermagnetic contact. We corroborate the former statement by using a rigorous functional-integral approach in Sec.~\ref{sec:path}. Therefore, this immediately excludes rough interfaces that do not conserve any of the momentum components. On the other hand, by introducing a preferred direction, a planar interface allows one to leverage the momentum-dependent spin-splitting of altermagnets and, as a result, obtain a thermoelectric response similar to that in the FM-SC-FM heterostructure, see Sec.~\ref{sec:model-FM}. The asymmetry between the spin-resolved spectral functions and, as a result, the thermoelectric effect is maximal when the altermagnetic lobes are perpendicular to the interface, see, e.g., Fig.~\ref{fig:model-setup}. If one of the lobes is rotated by $\pi/4$, the planar interface no longer induces the asymmetry between the spin species, hence, no thermoelectric effect is observed. This strong dependence on the crystallographic orientations of the altermagnets in the junction is a hallmark feature of altermagnets.

To support our qualitative picture, we calculate the thermoelectric response in an effective model of AM-SC bilayer coupled to another altermagnet. We use Eq.~(\ref{toy-model-momentum-Lij-def}) with the Green's functions (\ref{toy-model-G-R}) and (\ref{toy-model-G-L}). Integrating over the momentum component $k_{\perp}$ in Eq.~(\ref{toy-model-momentum-Lij-def}), we obtain
\begin{widetext}
\begin{eqnarray}
\label{toy-model-AM-Lij}
L_{ij} &=& 2\pi^2 \sqrt{2m}  L^3 \nu_0\int_{-\infty}^{\infty} d\omega  \int_{-\infty}^{\infty} d\xi_p \int_0^{2\pi}\frac{d\varphi}{2\pi} \sum_{\eta,s} \eta [t_s^{(R)}(\varphi)]^2 \frac{\Theta{\left((1 +\eta st_1^{(R)})(\eta \omega +\mu) -(\xi_p+\mu)\sin^2{\varphi}\left[1 - (t_1^{(R)})^2-(t_2^{(R)})^2\right]\right)}}{ \sqrt{(1 +\eta st_1^{(R)})(\eta \omega  +\mu) -(\xi_p+\mu)\sin^2{\varphi} \left[1 - (t_1^{(R)})^2-(t_2^{(R)})^2\right]}}
\nonumber\\
&\times&\sign{\omega -s J_L(\varphi)(\xi_p +\mu)} \left\{\eta \left[\omega -s J_L(\varphi)(\xi_p +\mu)\right] +\xi_p\right\} \df{\left[\omega -s J_L(\varphi) (\xi_p +\mu)\right]^2 -\xi_p^2 -\Delta^2} \frac{F_{ij;\eta}(\omega)}{4T\cosh^2{\left(\frac{\omega}{2T}\right)}},\nonumber\\
\end{eqnarray}
\end{widetext}
Due to the $\delta$-function, the integral over $\xi_p$ can be straightforwardly taken. The resulting expressions are cumbersome, hence, we do not present them in the main text.

In the case of the normal tunneling, i.e., with $\varphi^*\to0$, the expression for the Onsager coefficients (\ref{toy-model-AM-Lij}) simplifies as
\begin{widetext}
\begin{eqnarray}
\label{toy-model-AM-L11-ii-normal}
L_{ij}&\stackrel{\varphi^*\to0}{\approx}& \pi^2 \sqrt{\frac{2m}{\mu}} L^3 \nu_0 \int_{-\infty}^{\infty} d\omega \sum_{\eta,s} \frac{\eta [t_s^{(R)}(0)]^2}{\sqrt{1 +\eta st_1^{(R)}}} \sum_{\pm} \left\{\eta \left[\omega -s\mu J_L(0)\right] \pm \sqrt{\left[\omega -s\mu J_L(0)\right]^2 -\left[1 -J_L^2(0)\right]\Delta^2 } \right\} \nonumber\\
&\times&\frac{1}{1+\eta s J_L(0)} \frac{\sign{\omega -s\mu J_L(0) \mp sJ_L(0) \sqrt{\left[\omega -s\mu \Delta J_L(0)\right]^2 -\left[1 -J_L^2(0)\right] \Delta^2}} }{\sqrt{\left[\omega -s\mu J_L(0)\right]^2 -\left[1 -J_L^2(0)\right]\Delta^2}}
\frac{F_{ij;\eta}(\omega)}{4T\cosh^2{\left(\frac{\omega}{2T}\right)}},
\end{eqnarray}
\end{widetext}
where we also expanded in the large $\mu/T$ in the Green's function of the altermagnetic contact. As one can see from the above expression, the particle-hole symmetry breakdown for each of the spin species in the superconducting part of the junction is realized if the altermagnetism is weak, $\mu J_L(0) \sim \Delta$. Such a scenario is also presumably the most realistic in terms of permitting an altermagnetic spin-splitting coexisting with superconductivity due to the proximity effect~\cite{giil_arxiv_23}.

\begin{figure}[!ht]
\centering
\subfigure[]{\includegraphics[width=0.42\textwidth]{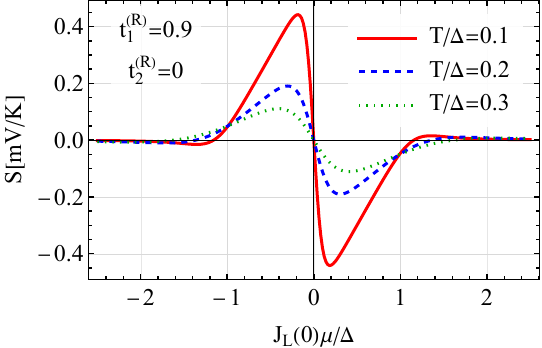}}
\subfigure[]{\includegraphics[width=0.42\textwidth]{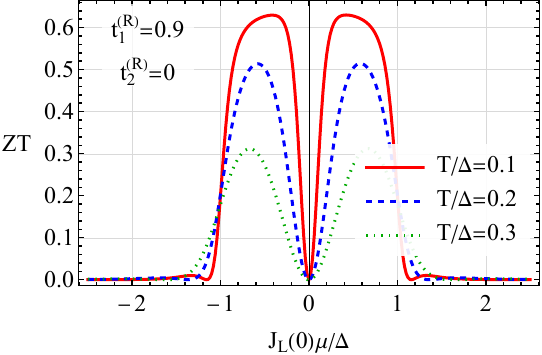}}
\caption{
The Seebeck coefficient $S$ (panel (a)) and the figure of merit $ZT$ (panel (b)) as a function of the altermagnetic strength $J_L(0)$ in the AM-SC-AM heterostructure for a few values of $T$. The altermagnetic parameters in the right altermagnet are $t_1^{(R)}=0.9$ and $t_2^{(R)}=0$, we disregard the spin-dependence of the tunneling coefficient, and we assume normal tunneling. See Eq.~(\ref{toy-model-AM-L11-ii-normal}) for the definition of the transport coefficients.
}
\label{fig:toy-model-AM-S-ZT}
\end{figure}

We present the Seebeck coefficient and the figure of merit for the AM-SC-AM heterostructure in Fig.~\ref{fig:toy-model-AM-S-ZT}. Comparing Figs.~\ref{fig:toy-model-FM-S-ZT} and \ref{fig:toy-model-AM-S-ZT}, we notice a similar shape of the curves as well as the suppression with temperature. The most drastic difference is in the scale of the altermagnetic strength in the AM-SC bilayer, which is determined by a parametrically small quantity $\Delta/\mu$. The magnitude of $S$ and $ZT$ in the AM-SC-AM heterostructure can be further enhanced by taking a stronger altermagnet in the right junction; this is similar to taking a stronger spin filtering coefficient $P$ or spin splitting $h_R$ in the FM-SC-FM heterostructure. Depending on the details of the tunneling, the inclusion of an explicit spin dependence in the tunneling coefficient may reduce or enhance the thermoelectric response. Since the spin-dependent tunneling coefficient is not crucial, we leave its discussion to a separate study. As with the ferromagnetic heterostructure, a wide tunneling cone (large $\varphi^*$ in Eq.~(\ref{model-setup-t-phi})) suppresses the thermoelectric response. Due to the interplay of the angular-dependent terms in the spectral functions, the dependence on $\varphi^*$ is nonmonotonic.

\section{Inverse proximity effect}
\label{sec:path}

In this Section, we use a different approach to the inverse proximity effect. Instead of the phenomenological momentum-dependent exchange field used in Sec.~\ref{sec:model}, we use the functional integral approach to derive the effective action of a superconductor proximitized with an altermagnet. The obtained action is used to calculate the corresponding Green's function in the AM-SC part of the junction; the transport coefficients straightforwardly follow from Eq.~(\ref{toy-model-momentum-Lij-def}).

\subsection{Functional integral approach}
\label{sec:path-action}

We start with the following action $S_{AM-S}$ describing an altermagnet, a superconductor, and tunneling between them:
\begin{eqnarray}
\label{path-action-S-def}
S_{AM-S} &=& S_S +S_{AM} +S_t,\\
\label{path-action-S-AM-def}
S_{AM} &=& -\frac{1}{L^d} \sum_{K} a_{K}^{\dag} \left[\omega -\xi_{\mathbf{k}} -\sigma_z J(\mathbf{k})\right] a_{K},\\
\label{path-action-S-S-def}
S_S &=& -\frac{1}{L^d} \sum_{K} c_{K}^{\dag} (\omega -\xi_{\mathbf{k}}) c_{K} \nonumber\\
&-& \frac{1}{L^{3d}} \sum_{K,K',Q} V_{K,K'} c_{K',\uparrow}^{\dag} c_{-K'+Q,\downarrow}^{\dag} c_{-K+Q,\downarrow}c_{K,\uparrow},\nonumber\\
\\
\label{path-action-S-t-def}
S_{t} &=& -\frac{1}{L^{2d}} \sum_{K,P} t_{K,P}\left(c_{K}^{\dag}a_{P} +a_{P}^{\dag} c_{K}\right),
\end{eqnarray}
where $a_{K} = \{a_{K,\uparrow}, a_{K,\downarrow}\}$, $c_{K} = \{c_{K,\uparrow}, c_{K,\downarrow}\}$, $K=\{\omega, \mathbf{k}\}$, and $\sum_{K} = \int d\omega/(2\pi) \sum_{\mathbf{k}}$. With the exception of altermagnetic exchange field $J(\mathbf{k})$ in Eq.~(\ref{path-action-S-AM-def}), the above action is similar to that used in Ref.~\cite{Hugdal-Sudbo:2019}.

The partition function is
\begin{equation}
\label{path-action-Z}
Z = \int Dc^{\dag}Dc Da^{\dag} Da \, e^{-S_S-S_{AM}-S_{t}}.
\end{equation}
Integrating out fermions in the altermagnet and ignoring the prefactor, we obtain
\begin{equation}
\label{path-action-Z-eff}
Z_{eff} = \int Dc^{\dag}Dc\, e^{-S_S} e^{\frac{1}{L^{2d}} \sum_{P,K}|t_{PK}|^2 c_{P}^{\dag} G_{AM,0}(K) c_{P}},
\end{equation}
where $G_{AM,0}(K) = \left[\omega -\xi_{\mathbf{k}} -\sigma_z J(\mathbf{k})\right]^{-1}$ is the Green's function of an altermagnetic layer in the absence of the proximity effect.

Therefore, the effective action of the proximitized superconductor reads
\begin{eqnarray}
\label{path-action-S-eff}
S_{eff} &=& -\frac{1}{L^d} \sum_{K,P} c^{\dag}_{K} \left[\delta_{K,P}G_{S,0}^{-1}(K) -|t_{KP}|^2 G_{AM,0}(P) \right]c_{K} \nonumber\\
&-& \frac{1}{L^{3d}} \sum_{K,K',Q} V_{K,K'} c_{K',\uparrow}^{\dag} c_{-K'+Q,\downarrow}^{\dag} c_{-K+Q,\downarrow}c_{K,\uparrow},
\end{eqnarray}
where $G_{S,0}^{-1}(K) = \omega -\xi_{\mathbf{k}}$. This action allows us to introduce the effective inverse Green's function as
\begin{equation}
\label{path-action-G-1}
G_{eff}^{-1}(K) = G_{S,0}^{-1}(K) - \Sigma_{AM}(K)
\end{equation}
with
\begin{equation}
\label{path-action-Sigma-1}
\Sigma_{AM}(K)  = \sum_{P} |t_{K,P}|^2 \frac{1}{\omega -\xi_{\mathbf{p}} -\sigma_z J(\mathbf{p})}
\end{equation}
being the self-energy due to the coupling to an altermagnet.

Let us proceed to the pairing term, i.e., the last term in the effective action (\ref{path-action-S-eff}). By using the standard Hubbard-Stratonovich transformation and the factorized form of the interaction $V_{K,K'} = g v(K)v(K')$, we obtain the following effective action:
\begin{eqnarray}
\label{path-action-S-eff-1}
S_{eff} &=& -\frac{1}{L^d} \sum_{K} c^{\dag}_{K} G_{eff}^{-1}(K) c_{K} +\frac{L^d}{g} \sum_{Q} \phi^{\dag}_{Q}\phi_{Q} \nonumber\\
&+& \frac{1}{L^d}\sum_{K,Q} v(K) \left[\phi_{Q}^{\dag} c_{K,\uparrow} c_{-K'+Q,\downarrow} +h.c.\right].
\end{eqnarray}

The effective action acquires a compact form in the Nambu space
\begin{equation}
\label{path-action-S-eff-2}
S_{eff} = -\frac{1}{2L^d} \sum_{K,K'} \Psi^{\dag}_{K} \hat{G}^{-1}(K,K') \Psi_{K'},
\end{equation}
where
\begin{equation}
\label{path-action-Nambu}
\Psi_{K} = \left\{c_{K,\uparrow}, c_{K,\downarrow}, c_{-K,\downarrow}^{\dag}, -c_{\uparrow,-K}^{\dag}\right\}^{T}
\end{equation}
and
\begin{equation}
\label{path-action-G-Nambu}
\hat{G}^{-1}(K,K') =
\begin{pmatrix}
\delta_{K,K'} G_{eff}^{-1}(K) & v(K) \phi_{K'-K}
\\
v(K) \mathcal{T} \phi_{K'-K} \mathcal{T}^{-1}
& -\delta_{K,K'} \mathcal{T} G_{eff}^{-1}(K) \mathcal{T}^{-1}
\end{pmatrix}
\end{equation}
is the inverse Nambu-Gorkov propagator. Here, $\mathcal{T} = i\sigma_y \mathcal{K} \Pi_{\mathbf{k} \to -\mathbf{k}}$ is the time-reversal symmetry operator and $\mathcal{K}$ is the complex conjugation operator.

In the mean-field approximation, we assume a uniform bosonic field $\phi_Q = \delta_{Q,0} \phi_{0}$ and define the superconducting order parameter as $\Delta(\mathbf{k}) = \phi_0 v(\mathbf{k})$. Then, the inverse Nambu-Gorkov propagator (\ref{path-action-G-Nambu}) reads
\begin{equation}
\label{path-action-G-Nambu-MF}
\hat{G}^{-1}(K,K') = \delta_{K,K'}
\begin{pmatrix}
G_{eff}^{-1}(K)  & \Delta(\mathbf{k}) \\
\Delta^{\dag}(\mathbf{k}) & - \mathcal{T} G_{eff}^{-1}(K) \mathcal{T}^{-1}
\end{pmatrix}.
\end{equation}
In what follows, we assume pairing in the spin-singlet channel, $\Delta(\mathbf{k})\propto \mathbb{I}_2$.

By inverting the matrix in Eq.~(\ref{path-action-G-Nambu-MF}), we find the following expression for the diagonal components of the Nambu-Gorkov function~\footnote{Off-diagonal components of the Nambu-Gorkov propagator or Gorkov functions, are not required for the tunneling current if the Andreev processes are neglected.}
\begin{equation}
\label{path-action-G-full}
G_{\eta,s}(\omega,\mathbf{p}) = \eta \frac{\tilde{\omega}_{\eta,p,s} +\tilde{\xi}_{\omega,p,s}}{\tilde{\omega}_{\eta,p,s}^2 -\tilde{\xi}_{\omega,p,s}^2 -|\Delta(\mathbf{p})|^2}
\end{equation}
with
\begin{eqnarray}
\label{path-action-teps-def}
\tilde{\omega}_{\eta,p,s} &=& \eta \omega +\frac{\Sigma_{-\eta,s}(\omega,\mathbf{p}) -\Sigma_{\eta,s}(\omega,\mathbf{p})}{2},\\
\label{path-action-txi-def}
\tilde{\xi}_{\omega,p,s} &=& \xi_{p} +\frac{\Sigma_{-\eta,s}(\omega,\mathbf{p}) +\Sigma_{\eta,s}(\omega,\mathbf{p})}{2},\\
\label{path-action-Sigma-def}
\Sigma_{\eta,s}(\omega,\mathbf{p}) &=& \sum_{P} |t_{K,P}|^2 \frac{1}{\eta \omega -\xi_{\eta \mathbf{p}} -s \eta J(\eta\mathbf{p})}.
\end{eqnarray}

The obtained Green's function (\ref{path-action-G-full}) is used in Eqs.~(\ref{model-setup-Is}) and (\ref{toy-model-momentum-Lij-def}) as $G_{L;\eta,s}(\omega,\mathbf{p})$; in addition, we replace $J(\eta\mathbf{p}) \to J_L(\eta\mathbf{p})$ and $|t_{K,P}|^2 \to |t_{K,P}^{(L)}|^2$.

\subsection{Inverse proximity effect in bilayer}
\label{sec:path-bilayer}

Before proceeding to the thermoelectric response, let us discuss the spectral properties of the AM-SC bilayer. In a bilayer, the tunneling coefficient preserves all components of the in-plane momenta, $|t_{K,P}|^2= \delta_{K,P} |t|^2$. This allows us to straightforwardly perform the summation over momenta in the self-energy (\ref{path-action-Sigma-def}) and obtain the following expressions for the parameters $\tilde{\omega}_{\eta,p,s}$ and $\tilde{\xi}_{\omega,p,s}$:
\begin{eqnarray}
\label{path-bilayer-teps}
\tilde{\omega}_{\eta,p,s} &=& \eta \omega -\eta |t|^2 \frac{\omega -s J(\mathbf{p})}{\left[\omega -s J(\mathbf{p})\right]^2 -\xi_p^2},\\
\label{path-bilayer-txi}
\tilde{\xi}_{\omega,p,s} &=& \xi_{p} + |t|^2 \frac{\xi_p}{\left[\omega -s J(\mathbf{p})\right]^2 -\xi_p^2},
\end{eqnarray}
where we also used the fact that in inversion-symmetric systems, $J(\eta\mathbf{p}) =J(\mathbf{p})$ and $\xi_{\eta \mathbf{p}} = \xi_{\mathbf{p}}$. The self-energy strongly affects the structure of the poles of the Nambu-Gorkov function (\ref{path-action-G-full}) and, as a result, the spectral properties of the bilayer.

\subsubsection{NM-SC bilayer}
\label{sec:path-bilayer-NM-SC}

In the absence of altermagnetism, i.e., for normal metal (NM)-superconductor bilayer, and assuming an $s$-wave superconducting gap, the four zeroes of the denominator of Eq.~(\ref{path-action-G-full}) can be straightforwardly found:
\begin{equation}
\label{path-bilayer-xi-poles}
\xi_{p,\pm}^2 = |t|^2 +\omega^2 -\frac{\Delta^2}{2} \pm \sqrt{4|t|^2 \left(\omega^2 -\frac{\Delta^2}{4}\right) +\frac{\Delta^4}{4}}.
\end{equation}
The gap is realized when all solutions $\xi_{p,\pm}$ have imaginary parts and, hence, they do not contribute to the DOS
\begin{equation}
\label{path-bilayer-DOS-def}
\nu_s(\omega) = \int \frac{d\mathbf{p}}{(2\pi)^2} A_{s}(\omega,\mathbf{p}),
\end{equation}
where
\begin{equation}
\label{path-bilayer-A-def}
A_{s}(\omega,\mathbf{p}) = -\frac{1}{2\pi} \sum_{\eta=\pm} \IM{G_{\eta,s}(\omega,\mathbf{p})}
\end{equation}
is the spin-resolved spectral function. This is the case for $|\omega| \leq \Delta_{\rm min}$, where the minigap $\Delta_{\rm min}$ reads
\begin{eqnarray}
\label{path-bilayer-minigap}
\Delta_{\rm min} &=& \frac{\Delta}{2}\left(\sqrt{1+\frac{4|t|^2}{\Delta^2}}-1\right) \,\Theta{\left(\frac{1+\sqrt{5}}{8} -\frac{|t|^2}{\Delta^2}\right)} \nonumber\\
&+& \frac{\Delta^2}{4|t|} \sqrt{\frac{4|t|^2}{\Delta^2}-1} \, \Theta{\left(\frac{|t|^2}{\Delta^2} -\frac{1+\sqrt{5}}{8}\right)}.
\end{eqnarray}
At strong tunneling, $|t|/\Delta \to \infty$, the minigap saturates to $\Delta/2$. For intermediate values of energy, $\Delta\left(\sqrt{1+4|t|^2/\Delta^2}-1\right)/2 <|\omega| < \Delta\left(\sqrt{1+4|t|^2/\Delta^2}+1\right)/2$, only half of the poles contribute leading to smaller DOS compared to the case $|t|=0$ and $|\omega|\gtrsim \Delta$.

We present the DOS for the NM-SC bilayer in Fig.~\ref{fig:path-bilayer-AM-SC}(a). As one can see, nonzero interlayer tunneling leads to the formation of the minigap $\Delta_{\rm min}<\Delta$ in the superconducting region, a feature which seems to often be overlooked in the literature. The minigap rises with the tunneling strength in agreement with Eq.~(\ref{path-bilayer-minigap}). Furthermore, we observe the splitting of the coherence peaks~\footnote{The structure of the peaks in the bilayer resembles that of a single-layer superconductor with an exchange field; the DOS in the bilayer is, however, spin-degenerate.}, which agrees with the four poles of Green's function, see Eq.~(\ref{path-bilayer-xi-poles}). These results for the NM-SC bilayer can also be deduced from the expressions of Ref.~\cite{Bergeret-Martin-Rodero:2005}.

\subsubsection{AM-SC bilayer}
\label{sec:path-bilayer-AM-SC}

\begin{figure}[!ht]
\centering
\subfigure[]{\includegraphics[width=0.42\textwidth]{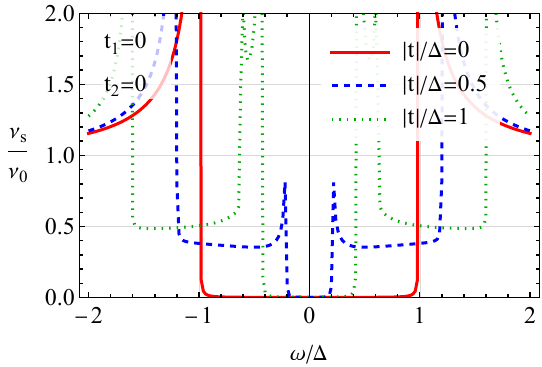}}
\subfigure[]{\includegraphics[width=0.42\textwidth]{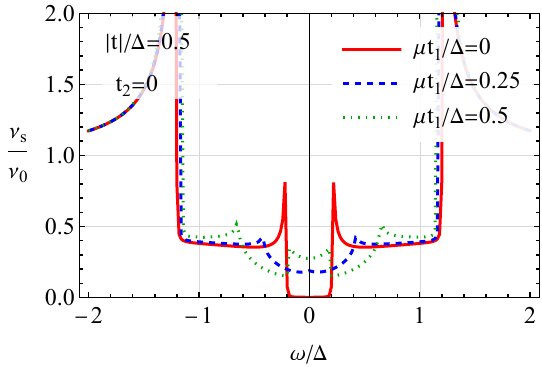}}
\caption{
The DOS in the superconducting part of the NM-SC bilayer ($t_1=t_2=0$) at a few values of $|t|$ (panel (a)) and in the superconducting part of the AM-SC bilayer at a few values of $t_1$ for $|t|/\Delta=0.5$ and $t_2=0$ (panel (b)). We use $J(\mathbf{p}) = (\xi_p+\mu)\left[t_1 \cos{(2\varphi)} +t_2 \sin{(2\varphi)}\right]$ and $\mu/\Delta=10^3$.
}
\label{fig:path-bilayer-AM-SC}
\end{figure}

Due to the momentum-dependence of the altermagnetic parameter $J(\mathbf{p})$ in the self-energy (\ref{path-action-Sigma-def}), the analytical analysis becomes more involved. As follows from our numerical results shown in Fig.~\ref{fig:path-bilayer-AM-SC}(b), the altermagnetic coupling closes the minigap. Similar to the thermoelectric transport in the effective model, see Sec.~\ref{sec:model-AM}, the effects of the altermagnetism are well-manifested only for small altermagnetic strengths, $\mu t_{1,2}/\Delta \sim 1$. For larger values of  $\mu t_{1,2}/\Delta$, the role of the self-energy diminishes~\footnote{The suppression of the inverse-proximity effect for $\mu t_{1,2}/\Delta \gg 1$ follows from the mismatch of the poles of the single-layer Green's function and the self-energy.} leading to the suppression of the inverse proximity effect. In the bulk of the bilayer, the orientation of the spin-split lobes of the altermagnet plays no role in the DOS which remains spin-degenerate.

\begin{figure}[!ht]
\centering
\subfigure[]{\includegraphics[width=0.42\textwidth]{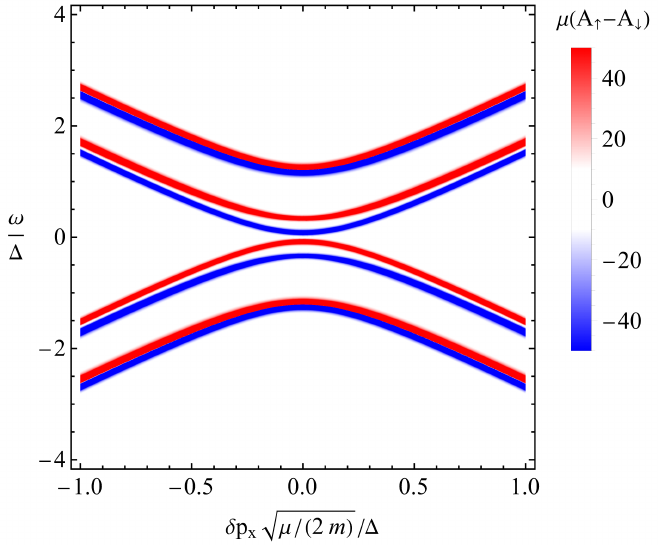}}
\subfigure[]{\includegraphics[width=0.42\textwidth]{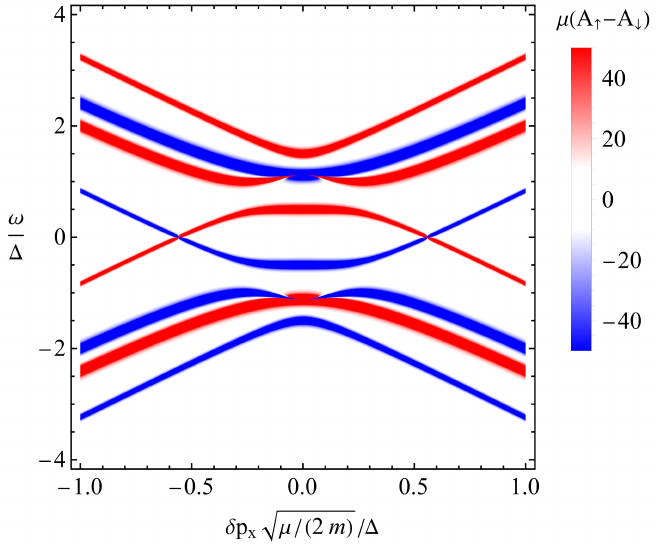}}
\caption{
The difference between the spin-up and spin-down spectral functions, $A_{\uparrow}(\omega,p_x,0)-A_{\downarrow}(\omega,p_x,0)$ as a function of the deviations from the Fermi momentum $\delta p_x = p_F -p_x$. Panels (a) and (b) correspond to $t_1=0.075\times \Delta/\mu$ and $t_1=0.5\times \Delta/\mu$, respectively.
We use $|t|/\Delta=0.5$, $t_2=0$, and $\mu/\Delta=10^3$.
}
\label{fig:path-bilayer-AM-SC-A}
\end{figure}

To show that the inverse proximity effect induces the momentum-dependent spin splitting in the superconductor, we present the spin-asymmetry of the spectral functions $A_{\uparrow}(\omega,p_x,0)-A_{\downarrow}(\omega,p_x,0)$ in Fig.~\ref{fig:path-bilayer-AM-SC-A}. As one can see, there is a noticeable momentum-dependent spin-splitting. The structure of the splitting is nontrivial for small values of the altermagnetic strength $t_1 \sim \Delta/\mu$, see Fig.~\ref{fig:path-bilayer-AM-SC-A}(b). A more detailed evolution of the spin-splitting with the altermagnetic strength is shown in Fig.~\ref{fig:app-1-AM-SC-A}, see Appendix~\ref{sec:App-1}.

We cross-verify the spectral properties of NM-SC and AM-SC bilayers in a lattice model, see Appendix~\ref{sec:App-2} for the details of the model. The key features of the spectral properties of the NM-SC and AM-SC bilayers are the same in the functional-integral approach and in a tight-binding lattice model, cf. Figs.~\ref{fig:path-bilayer-AM-SC}(a) and \ref{fig: App-2-2-B} as well as Figs.~\ref{fig:path-bilayer-AM-SC-A} and \ref{fig: App-2-2-A}.

\subsection{Seebeck coefficient and figure of merit}
\label{sec:path-S-ZT}

\begin{figure}[!ht]
\centering
\subfigure[]{\includegraphics[width=0.42\textwidth]{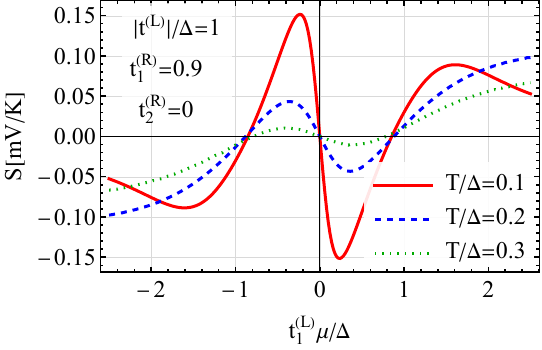}}
\subfigure[]{\includegraphics[width=0.42\textwidth]{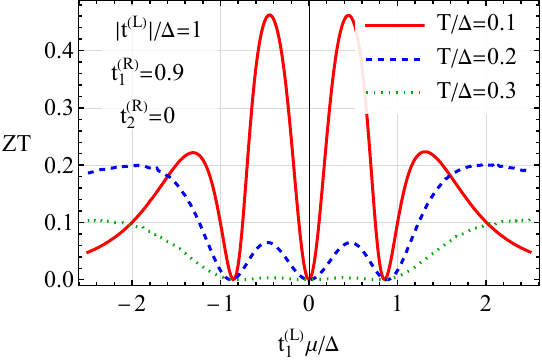}}
\caption{
The Seebeck coefficient $S$ (panel (a)) and the figure of merit $ZT$ (panel (b)) as a function of the altermagnetic parameter $t_1^{(L)}$ for a few values of $T$. We use $t_1^{(R)}=0.9$, $t_2^{(R)}=0$, $|t^{(L)}|/\Delta = 1$, and $\mu/\Delta=10^3$.
}
\label{fig:path-S-ZT}
\end{figure}

By using the formalism developed in Sec.~\ref{sec:path-action}, we calculate the thermoelectric response of the AM-SC-AM heterostructure. In view of a complicated structure of the full Green's function (\ref{path-action-G-full}), see also Eqs.~(\ref{path-action-teps-def})--(\ref{path-action-Sigma-def}), we focus on numerical results. We present the Seebeck coefficient $S$ and the figure of merit $ZT$ in Fig.~\ref{fig:path-S-ZT}. As in the effective model, see Fig.~\ref{fig:toy-model-AM-S-ZT}, the Seebeck coefficient and the figure of merit are peaked for small values of the altermagnetic strength $t_1^{(L)}\sim \Delta/\mu$. The double-peak structure of $ZT$ is more pronounced in the rigorous model compared to the effective model with the second peaks (i.e., at $|t_1^{(L)}| \gtrsim \Delta/\mu$) being of the same order of magnitude as those at $|t_1^{(L)}| \lesssim \Delta/\mu$; cf. Figs.~\ref{fig:toy-model-AM-S-ZT}(b) and \ref{fig:path-S-ZT}(b).

\begin{figure}[!ht]
\centering
\subfigure[]{\includegraphics[width=0.42\textwidth]{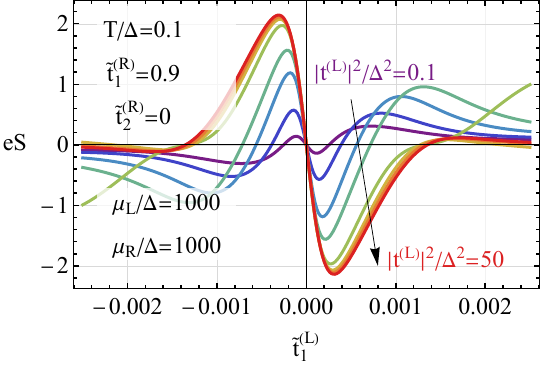}}
\subfigure[]{\includegraphics[width=0.42\textwidth]{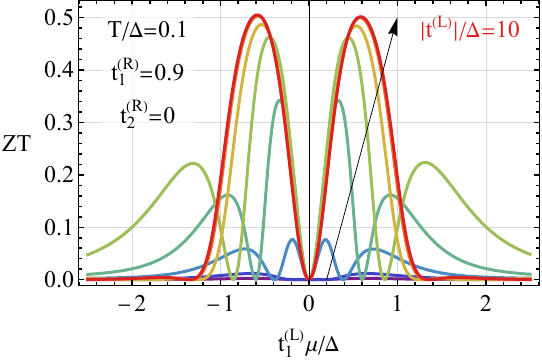}}
\caption{
The dependence of the Seebeck coefficient $S$ (panel (a)) and the figure of merit $ZT$ (panel (b)) on the altermagnetic strength at a few values of the inter-layer tunneling strength $|t^{(L)}|$. Colored lines correspond to $|t^{(L)}|/\Delta=\left\{0.1,0.25,0.5,0.75,1,2,10\right\}$. We use $t_1^{(R)}=0.9$, $t_2^{(R)}=0$, and $\mu/\Delta=10^3$.
}
\label{fig:interface-full-ZT-delta-t2L}
\end{figure}

The treatment of the inverse proximity effect based on the functional-integral approach introduces the dependence on the inter-layer tunneling constant $|t^{(L)}|$. As one can see from Fig.~\ref{fig:interface-full-ZT-delta-t2L}, larger values of the tunneling constant are beneficial for the thermoelectric response. However, the Seebeck coefficient and the figure of merit show the saturation behavior at $|t^{(L)}| \sim \Delta$.

As in the effective model, there is a strong dependence on the relative orientation of the spin-polarized lobes in the altermagnets. In presenting the results for the thermoelectric response, we focus on the case of maximal spin splitting where the altermagnetic lobes are normal to the interface both in the bilayer and the altermagnetic contact. Rotation away from this configuration decreases the effective spin splitting and, as a result, reduces the thermoelectric response.

Thus, as one can see comparing Figs.~\ref{fig:toy-model-AM-S-ZT} and \ref{fig:path-S-ZT}, the effective model and more rigorous treatment of the inverse proximity effect based on the functional integration agree well in the key aspects of the thermoelectric response such as the values of the altermagnetic splitting corresponding to the largest thermoelectric response.

\section{Discussion and Summary}
\label{sec:Summary}

In this paper, we investigated the thermoelectric response of the altermagnetic-superconductor heterostructures. We showed that altermagnets provide a viable way to achieve a sizable thermoelectric response that, unlike previously studied ferromagnet-superconductor heterostructures, is free of any magnetic fields. The latter property allows for better miniaturization, which is important for spintronic devices.

The key ingredients necessary for the thermoelectric effect include (i) the induced altermagnetic spin-splitting in the superconductor via the inverse proximity effect and (ii) the spin-selective tunneling between the proximitized superconductor and an electric contact, see Fig.~\ref{fig:model-setup}. The former allows for broken particle-hole symmetry for each of the spin species and the latter allows one to distinguish between the spin-split particles. We achieve the spin-splitting in a 2D superconductor by proximitizing it with a 2D altermagnet. The spin-selective tunneling is realized by coupling the obtained altermagnet-superconductor bilayer to another 2D altermagnet; the planar interface introduces the directional dependence and allows us to use the momentum-dependent spin splitting of altermagnets.
We underline that to observe this effect experimentally, it is not necessary to use an altermagnet close to an actual 2D limit (i.e., extremely thin films): we have performed the calculations in 2D for simplicity, as it is the smallest dimension that allows us to capture the altermagnetic characteristic features. The thermoelectric effects predicted here persist even in 3D.

In our modeling of the inverse proximity effect in the altermagnet-superconductor bilayer, we used an effective model with a momentum-dependent exchange field, see Eq.~(\ref{model-setup-H-AM-SC}) in Sec.~\ref{sec:model-setup}, and a more rigorous functional-integral approach, see Sec.~\ref{sec:path}. The simplicity of the effective model allows for reasonably compact and transparent expressions for the thermoelectric coefficients, see Eqs.~(\ref{toy-model-AM-Lij}) and (\ref{toy-model-AM-L11-ii-normal}). As a measure of the thermoelectric effect, we use the Seebeck coefficient $S$ and the figure of merit $ZT$ defined in Eqs.~(\ref{toy-model-momentum-S-def}) and (\ref{toy-model-momentum-ZT-def}), respectively. The corresponding results are shown in Fig.~\ref{fig:toy-model-AM-S-ZT}. As in the case of ferromagnetic heterostructures, see Sec.~\ref{sec:model-FM} and Fig.~\ref{fig:toy-model-FM-S-ZT}, the figure of merit is a nonmonotonic function of the altermagnetic field strength $J_L$ reaching its maximum, however, at parametrically small values $J_L \Delta/\mu \sim 1$. Another hallmark feature of altermagnets is the strong dependence on the orientation of crystallographic axes: the thermoelectric effects are maximal if one of the spin-polarized lobes of the altermagnetic Fermi surfaces is normal to the interface, see Fig.~\ref{fig:model-setup}. The thermoelectric response is reduced if the lobes are rotated away from this configuration.

The results of the effective model are in good qualitative agreement with those obtained via the functional-integral approach, see Secs.~\ref{sec:path-action} and \ref{sec:path-S-ZT}. Indeed, comparing Figs.~\ref{fig:toy-model-AM-S-ZT} and \ref{fig:path-S-ZT}, we note a similar shape and magnitude of $S$ and $ZT$; some finer features, such as the structure of the peaks, are, however, different. The stronger interlayer coupling is beneficial for the thermoelectric response, however, the Seebeck coefficient and the figure of merit saturate when the coupling is of the order of the superconducting gap. In all models, the thermoelectric response is enhanced at lower temperatures. The enhancement, however, is restricted by broadening effects, which are inevitably present in realistic materials, see Fig.~\ref{fig:toy-model-FM-S-ZT}(c). Quantitatively, we find that both $S$ and $ZT$ can reach useful magnitudes in the altermagnetic case, but both of these quantities are consistently smaller compared to what is obtained when using ferromagnets instead. Thus, the question of whether altermagnets or ferromagnets are most beneficial when it comes to thermoelectric effects with superconductors comes down to a trade-off: one can either get large thermoelectric effects, with the drawback of disturbing magnetic stray fields restricting miniaturization, or one can get moderate thermoelectric effects completely void of magnetic fields.

In addition to the thermoelectric response, we apply the functional-integral approach to study the spectral properties of altermagnet-superconductor bilayers. The results of Sec.~\ref{sec:path-bilayer} reveal the formation of the minigap in the DOS determined by the interlayer tunneling strength, see Eq.~(\ref{path-bilayer-minigap}) and Fig.~\ref{fig:path-bilayer-AM-SC}(a). As with the thermoelectric response, the altermagnetic spin-splitting has the most pronounced effect at small values of altermagnetic strength and leads to the closure of the minigap, see Fig.~\ref{fig:path-bilayer-AM-SC}(b). The spin-splitting induced by the inverse proximity effect in the superconducting part of the AM-SC bilayer can be probed via the spectral function, see Fig.~\ref{fig:path-bilayer-AM-SC-A}. The obtained results for the spectral properties are corroborated in a lattice model of the AM-SC bilayer, see Figs.~\ref{fig: App-2-2-B} and \ref{fig: App-2-2-A} in the appendix.

While in the present paper, we focused on the case of 2D altermagnets and superconductors, the obtained results could be straightforwardly extended to the case of 3D junctions and more complicated geometries of the contacts. For concreteness, we use altermagnets with a $d$-wave symmetry of the spin-split Fermi surfaces as a representative example. However, we expect qualitatively the same results for $g$- and $i$-wave altermagnets.

\begin{acknowledgments}
We acknowledge useful communications with Morten~Amundsen and Henning G.~Hugdal on the inverse proximity effect.
This work was supported by the Research Council of Norway through Grant No. 323766 and its Centres of Excellence funding scheme Grant No. 262633 “QuSpin.” Support from Sigma2 - the National Infrastructure for High-Performance Computing and Data Storage in Norway, project NN9577K, is acknowledged.
\end{acknowledgments}

\appendix

\begin{widetext}

\section{Spectral functions in AM-SC bilayer}
\label{sec:App-1}

In this Section, we present the spectral functions for a wider range of the inter-layer coupling parameters, see Fig.~\ref{fig:app-1-AM-SC-A}. Altermagnetism is manifested in a spin-splitting of energy levels, which is manifested in the difference between the spin-up and spin-down spectral functions. With the rise of the altermagnetic strength, the spin-split energy levels overlap and rearrange allowing for crossings at $\omega=0$, see Fig.~\ref{fig:app-1-AM-SC-A}(b). The intensity of these crossings diminishes at stronger altermagnetic parameters ultimately leaving only a single set of parabolic bands for the momenta near the Fermi level, see Figs.~\ref{fig:app-1-AM-SC-A}(e) and \ref{fig:app-1-AM-SC-A}(f).

\begin{figure*}[!ht]
\centering
\subfigure[$t_1 =0.05\times \Delta/\mu$]{\includegraphics[width=0.31\textwidth]{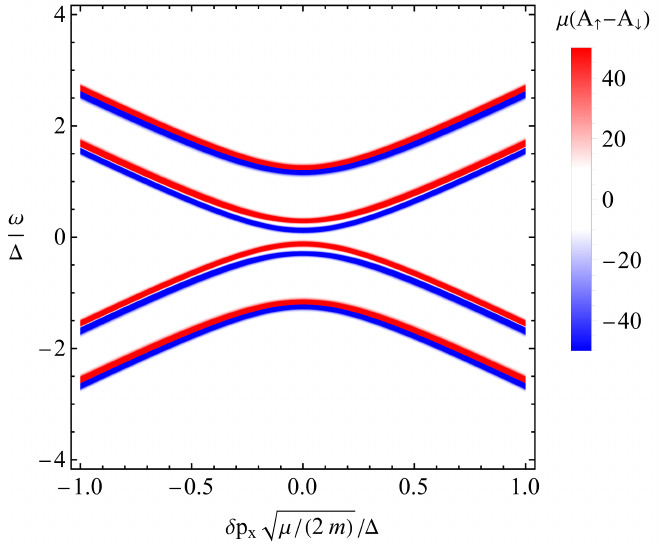}}
\subfigure[$t_1=0.25\times \Delta/\mu$]{\includegraphics[width=0.31\textwidth]{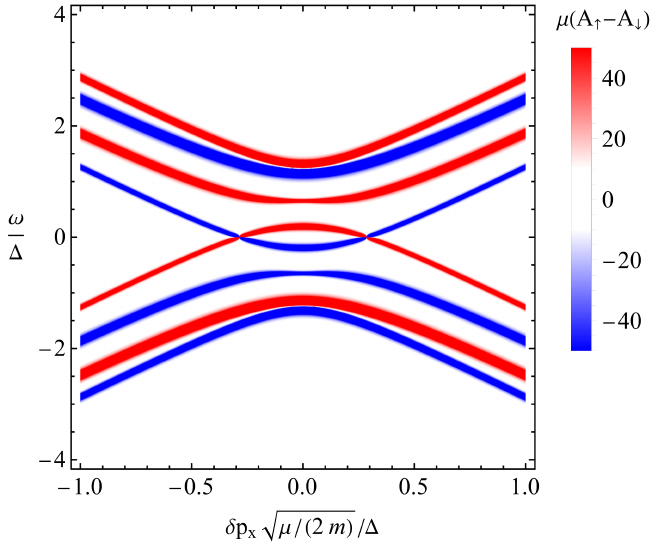}}
\subfigure[$t_1 =0.5\times \Delta/\mu$]{\includegraphics[width=0.31\textwidth]{Fig6b.pdf}}
\subfigure[$t_1 =\Delta/\mu$]{\includegraphics[width=0.31\textwidth]{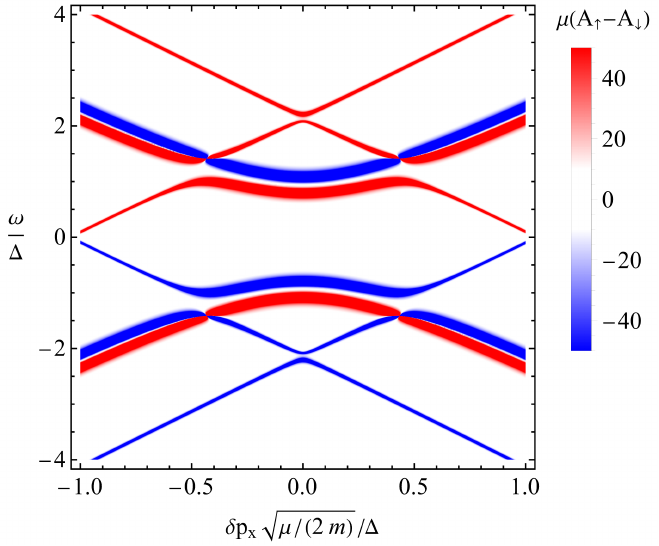}}
\subfigure[$t_1 =2\times \Delta/\mu$]{\includegraphics[width=0.31\textwidth]{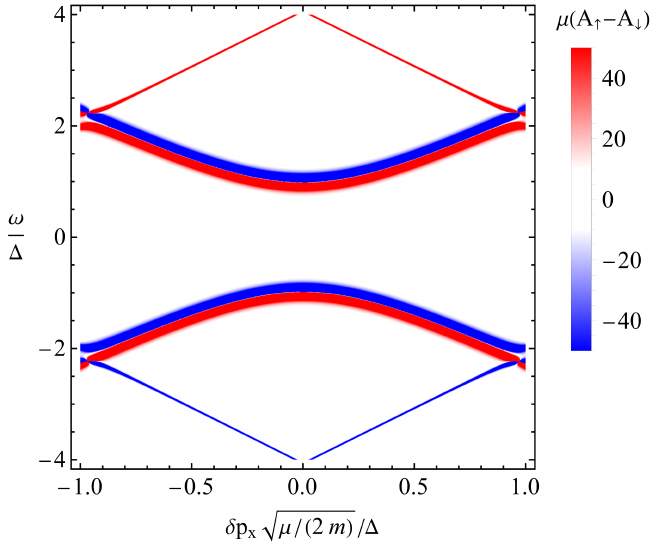}}
\subfigure[$t_1 =5\times \Delta/\mu$]{\includegraphics[width=0.31\textwidth]{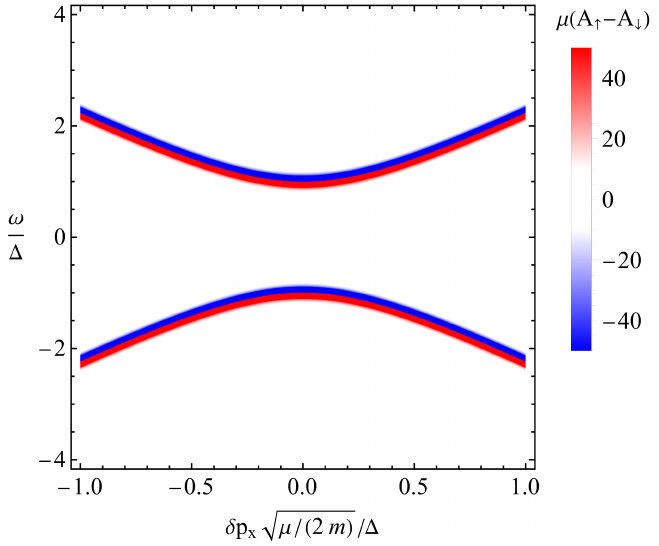}}
\caption{
The difference between the spin-up and spin-down spectral functions, $A_{\uparrow}(\omega,p_x,0)-A_{\downarrow}(\omega,p_x,0)$ as a function of the deviations from the Fermi momentum $\delta p_x = p_F -p_x$. In all panels, we use $|t|/\Delta=0.5$, $t_2=0$, and $\mu/\Delta=10^3$.
}
\label{fig:app-1-AM-SC-A}
\end{figure*}

\section{Lattice model}
\label{sec:App-2}

We consider a Bogoliubov-de Gennes lattice model given by the following Hamiltonian:
\begin{equation}
    H=\frac{1}{2}\sum_{p_x,p_y}\begin{pmatrix}
        \Tilde c_{p_x, p_y, \text{NM/AM}}^\dagger & \Tilde c_{p_x, p_y, \text{SC}}^\dagger
    \end{pmatrix}
    \begin{bmatrix}
        h_{\text{NM/AM}} & -\tau_{\text{layer}} \sigma_z \otimes\mathbb{I}_2 \\
        -\tau_{\text{layer}} \sigma_z \otimes\mathbb{I}_2 & h_{\text{SC}}
    \end{bmatrix}
    \begin{pmatrix}
        \Tilde c_{p_x, p_y, \text{NM/AM}}^{\vphantom{\dagger}} \\ \Tilde c_{p_x, p_y, \text{SC}}^{\vphantom{\dagger}}
    \end{pmatrix},
\end{equation}
where we have introduced the basis
\begin{equation}
        \Tilde c_{p_x, p_y, \text{layer}}^{\vphantom{\dagger}}
    = \begin{pmatrix}
         c_{p_x, p_y, \text{layer}, \uparrow}^{\vphantom{\dagger}} \\  c_{p_x, p_y, \text{layer}, \downarrow}^{\vphantom{\dagger}}
         \\ c_{-p_x, -p_y, \text{layer}, \uparrow}^{{\dagger}}
         \\ c_{-p_x, -p_y, \text{layer}, \downarrow}^{{\dagger}}
    \end{pmatrix}
\end{equation}
and the block elements
\begin{gather}
    h_{\text{NM/AM}}=[-2\tau(\cos p_x + \cos p_y ) - \mu] \sigma_z \otimes\mathbb{I}_2 -[2\tau_{1}(\cos p_x  - \cos p_y )]\sigma_z\otimes\sigma_z \\[10pt]
    h_{\text{SC}}= i(\Delta_{\text{layer}}\sigma^+ - \Delta_{\text{layer}}^*\sigma^-)\otimes\sigma_y.
\end{gather}
Here $\sigma^\pm = \sigma_x \pm i\sigma_y$, $\otimes$ is the Kronecker product, $\tau$ is the in-plane tight-binding hopping parameter, $\tau_{\text{layer}}$ couples the two layers in the bilayer, while $\tau_1$ determines the strength of the altermagnetic term. The gap in the bilayer structure is obtained self-consistently using the gap equation
\begin{equation}
    \Delta_{\text{layer}}=\frac{1}{N_xN_y}\sum_{p_x,p_y}U  \big\langle c_{p_x, p_y, \text{layer}, \uparrow}^{\vphantom{\dagger}} c_{-p_x, -p_y, \text{layer}, \downarrow}^{\vphantom{\dagger}} \big\rangle,
\end{equation}
expressed through the operators which diagonalize the Hamiltonian. The attractive interaction $U$ is taken to be non-zero only in the SC layer.

\subsection{Superconducting DOS and minigap formation}
\label{sec:App-2-1}

Using the model defined above, we consider a superconductor-normal metal bilayer. The DOS in the superconductor is shown in Fig.~\ref{fig: App-2-2-B} for an attractive strength $U=3\tau$ with a chemical potential $\mu=-2\tau$. The in-plane system dimension was $500\times500$ momentum modes.

In the limiting case of vanishing interlayer coupling ($\tau_\text{layer}=0$), the SC depicts a gap $\Delta_0\simeq 0.45\tau$. While the magnitude is unrealistic, it is chosen to make the appearance of the minigap feature more prominent. As the coupling between layers becomes non-zero, the SC gap is affected. We observe the emergence of the minigap structure inside the original SC gap, discussed in Sec.~\ref{sec:path-bilayer}.
In agreement with the analytical result in Eq.~(\ref{path-bilayer-minigap}), the inter-layer tunneling opens a minigap in the DOS that increases with the tunneling strength, cf. Figs.~\ref{fig:path-bilayer-AM-SC}(a) and \ref{fig: App-2-2-B}.

\begin{figure}
    \centering
    \includegraphics[width=0.50\linewidth]{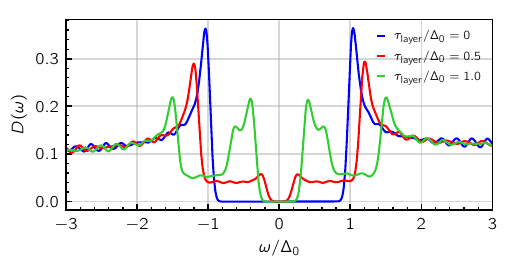}
    \caption{
    The DOS in the superconducting layer is shown for an SC-NM bilayer coupled by the interlayer tight-binding parameter $\tau_{\text{layer}}$, as a function of energy normalized by the gap $\Delta_0$. The attractive interaction in the SC was set to $U=2\tau$ and the chemical potential to $\mu=-2\tau$. In the absence of inter-layer coupling ($\tau_{\text{layer}}=0$), the SC gap is $\Delta_0\simeq0.1\tau$. As the layers are coupled, the minigap feature discussed in Sec.~\ref{sec:path-bilayer} arises inside the original gap.}
    \label{fig: App-2-2-B}
\end{figure}

\subsection{AM / SC bilayer and spin-split spectral function}
\label{sec:App-2-2}

We consider now an SC-AM bilayer with an aim to quantify the momentum-dependent spin-splitting induced by the AM in the SC layer. The difference between spin-up and spin-down spectral functions is shown in Fig.~\ref{fig: App-2-2-A}. We use $U=3\tau$, $\mu=-2\tau$, $\tau_\text{layer}=0.25\tau$ and various strengths of the AM parameter $\tau_1$. The system consists of $250\times250$ momentum modes. Note that the evolution of the spectral functions is similar in linearized and lattice models, cf. Figs.~\ref{fig:app-1-AM-SC-A} and \ref{fig: App-2-2-A}.

Thus, we showed that the key features of the spectral properties of the NM-SC and AM-SC bilayers are the same in the functional-integral approach and in a tight-binding lattice model.

\begin{figure*}
    \centering
    \subfigure[$\tau_1 = 0.05\tau$]{\includegraphics[width=0.31\textwidth]{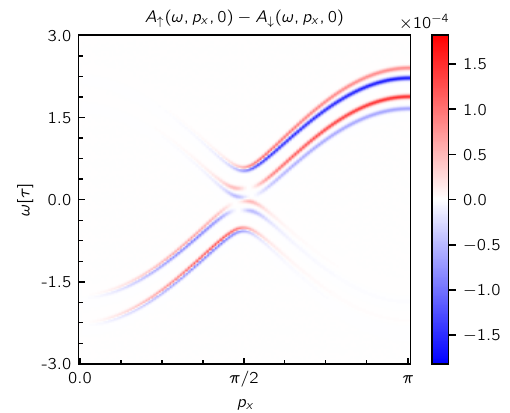}}
    \subfigure[$\tau_1 = 0.25\tau$]{\includegraphics[width=0.31\textwidth]{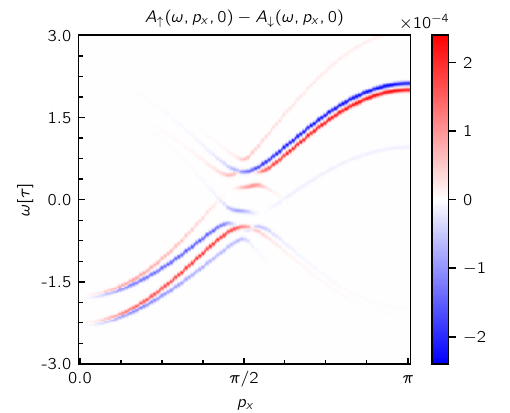}}
    \subfigure[$\tau_1 = 0.50\tau$]{\includegraphics[width=0.31\textwidth]{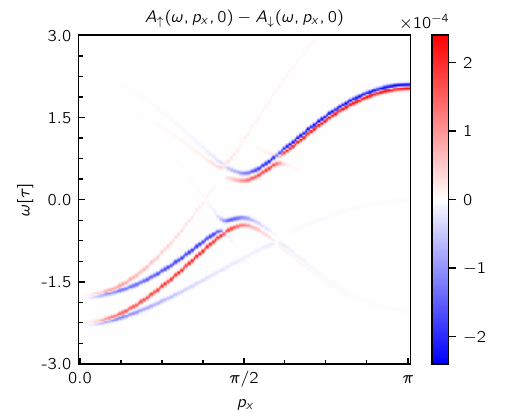}}
    \subfigure[$\tau_1 = 1.00\tau$]{\includegraphics[width=0.31\textwidth]{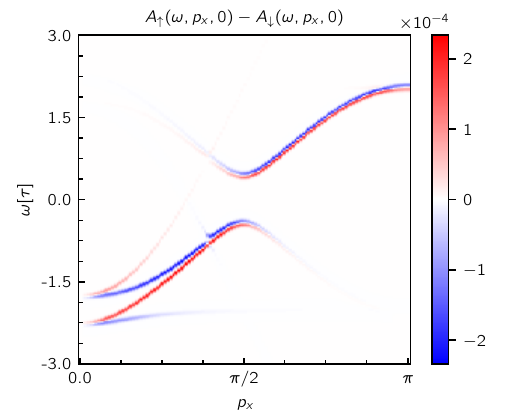}}
    \subfigure[$\tau_1 = 1.50\tau$]{\includegraphics[width=0.31\textwidth]{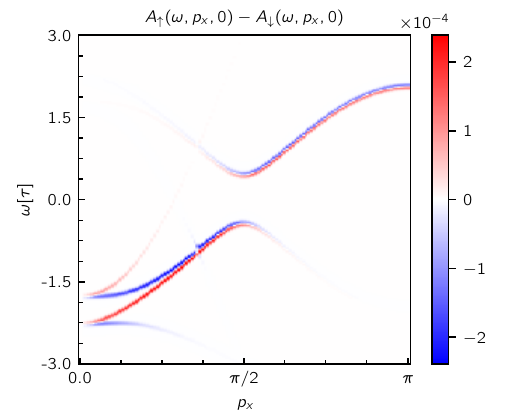}}
    \caption{
    The difference between the spin-up and spin-down spectral functions $A_{\uparrow}(\omega,p_x,0)-A_{\downarrow}(\omega,p_x,0)$ in the SC is shown for an SC-AM bilayer with $U=3\tau$, $\tau_{\text{layer}}=0.25\tau$, and $\mu=-2\tau$ for varying AM strengths $\tau_1$. We plot the $p_y=0$ slice, showing the spectral function as a function of $p_x$ and $\omega$. }
    \label{fig: App-2-2-A}
\end{figure*}
\end{widetext}

\bibliography{library-short}

\end{document}